\documentclass[num-refs,twocolumn]{wiley-article}

\usepackage{siunitx}
\usepackage[version=4]{mhchem}

\papertype{Original Article}
\paperfield{Journal Section}

\title{Site-selective and real-time observation of bimolecular electron transfer during photocatalytic water splitting}

\author[1, 2\authfn{1}]{Alexander Britz}
\author[3]{Sergey I. Bokarev}
\author[1,\authfn{2}]{Tadesse A. Assefa}
\author[4]{\'{E}va G. Bajn{\'{o}}czi}
\author[4]{Zolt{\'{a}}n N{\'{e}}meth}
\author[4]{Gy{\"{o}}rgy Vank{\'{o}}}
\author[5]{Nils Rockstroh}
\author[5]{Henrik Junge}
\author[5]{Matthias Beller}
\author[6]{Gilles Doumy}
\author[6]{Anne Marie March}
\author[6]{Stephen H. Southworth}
\author[3]{Stefan Lochbrunner}
\author[3]{Oliver Kühn}
\author[1,2]{Christian Bressler}
\author[1,7\authfn{1}\authfn{3}]{Wojciech Gawelda}

\affil[1]{European XFEL, Holzkoppel 4, 22869 Schenefeld, Germany}
\affil[2]{~The Hamburg Centre for Ultrafast Imaging, Luruper Chaussee 149, 22761 Hamburg, Germany}
\affil[3]{Institut f\"ur Physik, Universit\"at Rostock, Albert-Einstein-Str. 23-24, 18059 Rostock, Germany}
\affil[4]{Wigner Research Centre for Physics, H-1525 Budapest, Hungary}

\affil[5]{Leibniz-Institut f\"ur Katalyse an der Universit\"at Rostock, Albert-Einstein-Str. 29a, 18059 Rostock, Germany}

\affil[6]{Chemical Sciences and Engineering Division, Argonne National Laboratory, 9700 S. Cass Ave, 60439 Lemont, IL, USA}

\affil[7]{Faculty of Physics, Adam Mickiewicz University, ul. Uniwersytetu Pozna\'nskiego 2, Pozna\'n, 61-614, Poland}

\corraddress{Alexander Britz\authfn{1}, Wojciech Gawelda\authfn{1}}
\corremail{mail@alexbritz.de, wojciech.gawelda@amu.edu.pl}

\presentadd[\authfn{2}]{Condensed Matter Physics and Materials Science Department, Brookhaven National Laboratory, Upton, NY 11793, USA}
\presentadd[\authfn{3}]{Department of Chemistry, Faculty of Sciences, Universidad Aut\'onoma de Madrid and IMDEA-Nanoscience, Ciudad Universitaria de Cantoblanco, 28049 Madrid, Spain}

\runningauthor{A. Britz et al.}

\begin{document}

\begin{frontmatter}
\maketitle

\begin{abstract}
Time-resolved X-ray absorption spectroscopy has been utilized to monitor the bimolecular electron transfer in a photocatalytic water splitting system for the first time. This has been possible by uniting the local probe and element specific character of X-ray transitions with insights from high-level ab initio calculations. The specific target has been a heteroleptic \ce{[Ir$^{\textnormal{\tiny{III}}}$(ppy)2(bpy)]+} photosensitizer, in combination with triethylamine as a sacrificial reductant and \ce{Fe$_3$(CO)$_{12}$} as a water reduction catalyst. The relevant molecular transitions have been characterized via high-resolution Ir L-edge X-ray absorption spectroscopy on the picosecond time scale. The present findings enhance our understanding of functionally relevant bimolecular electron transfer reactions and thus will pave the road to rational optimization of photocatalytic performance.

\keywords{homogeneous catalysis, photocatalytic water splitting, electron transfer, x-ray absorption spectroscopy, high energy-resolution fluorescence detection XAS, ultrafast XAS}
\end{abstract}
\end{frontmatter}
\section{Introduction}
\begin{figure}[ht]
\centering
  \includegraphics[height=5cm]{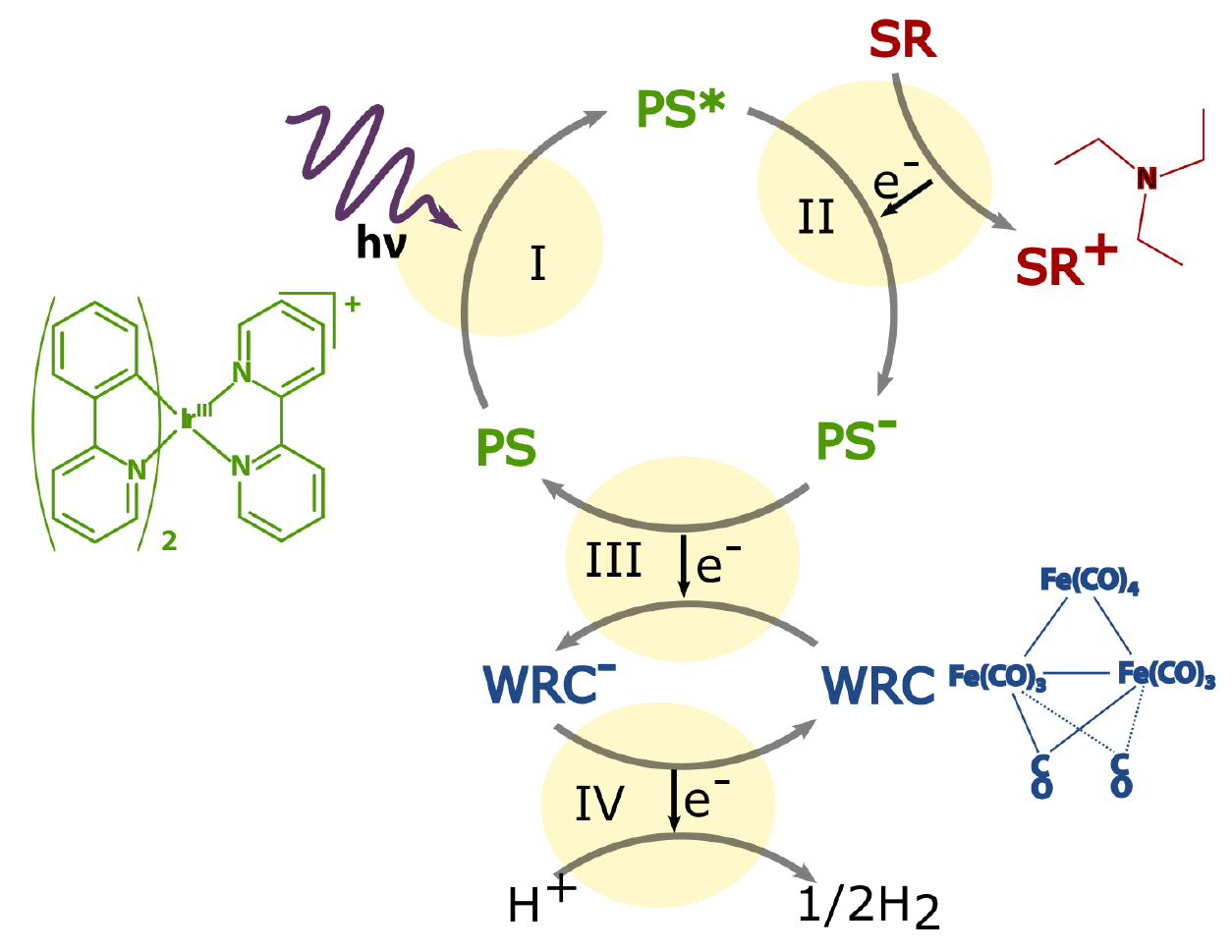}
  \caption{Reaction cycle of a photocatalytic system consisting of a homogeneous solution of a sacrificial reductant (SR), a photosensitizer (PS) and a water reduction catalyst (WRC) for solar hydrogen generation \cite{Gartner2009,Gartner2011, Gartner2012, Neubauer2014, Bokarev2015}. The system discussed here consists of the PS \ce{[Ir$^{\textnormal{\tiny{III}}}$(ppy)2(bpy)]+}, triethylamine (TEA) as the SR and \ce{Fe3(CO)12} as the WRC.}
  \label{fig_reactioncycle}
\end{figure}

Water splitting has gained increased scientific interest in the past years as a sustainable source of carbon-free fuels, which could become a future alternative to the existing fossil-based sources.\cite{Armaroli2007} In particular, solar light-induced catalytic splitting of water into hydrogen and oxygen is a very promising approach to generate hydrogen as a solar fuel, which undergoes combustion without producing undesired CO$_2$.\cite{Lewis2006} Applying catalysts allows to overcome the relatively high thermodynamical barriers associated with the chemical reaction steps leading to the evolution of oxygen and hydrogen.\cite{Sanderson2008} In particular, homogeneous catalysis is rapidly emerging in this field, and various protocols for the generation of oxygen\cite{Sala2009} and hydrogen\cite{Wang2009} from water have been published. A key challenge in this area is the design of more efficient photosensitizers (PS) that absorb well visible light leading to charge-separated and long-lived electronically excited states.\cite{Bokareva2017} The latter are typically associated with strong reductive and oxidative power and have to provide negative and positive charge with sufficient driving force for enabling water reduction (hydrogen generation) and oxidation (oxygen generation), respectively. Metalorganic complexes represent a very promising class of PSs since many of them feature rather long living metal-to-ligand charge transfer triplet ($^3$MLCT) states and, therefore, are currently subject of intense research activities. While at the beginning ruthenium complexes were mostly applied meanwhile, amongst others, iridium (Ir), copper, and iron complexes are broadly investigated.
Iridium complexes such as the heteroleptic complex \ce{[Ir$^{\textnormal{\tiny{III}}}$(ppy)2(bpy)]+} (where ppy = 2-phenylpyridine and bpy = 2,2'-bipyridine), turned out to be promising candidates for PS in water splitting.\cite{Goldsmith2005,Tinker2007}
G\"artner et al. reported that this IrPS results, in combination with iron carbonyl complexes as catalysts, in efficient homogenous systems for photocatalytic hydrogen generation.\cite{Gartner2012} 
Under optimized conditions exceptionally high internal quantum yields exceeding 40\% were achieved. \cite{gartner11_6998,junge17_14} The work on copper, iron, and other non-precious metals is driven by the goal to replace the rare noble metals by more abundant and less expensive elements. In the case of iron PSs the research is still focused on the design of complexes with suitable long-lived MLCT states. \cite{Wenger2019, Zimmer2018, Harlang2015} Promising results in photocatalytic water reduction were already reported for homogeneous photocatalytic systems based on Cu(I) PS complexes as well as on zinc porphyrins. \cite{Kuang2002, Lazarides2014, McCullough2018} However, all systems are still far from being applied in the real world due to their insufficient performance. Detailed mechanistic studies applying operando methods and stationary as well as time-resolved spectroscopy are performed in order to understand the individual reaction steps, the microscopic mechanisms, and performance limiting factors and to provide in this way guidelines for a better system design. \cite{Fischer2014,Arias-Rotondo2016,Papcke2020} Systems based on IrPSs are particularly well suited for such investigations since they show high yields and several spectroscopic experiments provide already a solid data base for the interpretation of results obtained by more sophisticated approaches. \cite{Gartner2009, junge17_14, Neubauer2014, Tschierlei2016}
\newline
Along this line we present the application of a suite of complementary time-resolved (TR) hard X-ray spectroscopies with picosecond (ps) temporal resolution, in combination with ab initio electronic structure calculations to study a fully functional photocatalytic system for solar hydrogen generation (Fig. \ref{fig_reactioncycle}). It includes triethylamine (TEA) acting as a sacrificial reductant (SR) that quenches the photoexcited \ce{[Ir$^{\textnormal{\tiny{III}}}$(ppy)2(bpy)](PF6)} and a \ce{Fe3(CO)12} complex as the water reduction catalyst.
 Our study aims to complete the current picture\cite{Neubauer2014,Bokarev2015} of the influence of geometric and/or electronic structures of long-lived ($>100$~ps) charge-separated excited states on the system's performance. This includes the efficiency of light-harvesting, subsequent reductive quenching of the Ir-based PS by the SR molecules, and the intermolecular electron transfer to the water reduction catalyst (WRC). A deeper understanding of the structure-function relationship is critical for the development and optimization of PS complexes for efficient hydrogen production in molecular photocatalysis. \newline
 The function of this system and its constituents has been described previously in Refs \cite{Gartner2009, Gartner2011,Bokarev2015,fischer16_404}, cf. Fig. \ref{fig_reactioncycle}. It involves a four-step process, in which the Ir-based PS absorbs a photon and is excited to an electronically excited state PS* (step I), which can be subsequently reduced by an electron transfer from a TEA molecule yielding PS$^{-}$ (step II). The next step involves another electron transfer to the WRC (step III), which then reduces a proton of a nearby \ce{H3O+} molecule to hydrogen (step IV).
The photocatalytic reaction steps reported here are initiated by excitation in the intense metal-to-ligand charge transfer (MLCT) bands in the near UV range around 350 nm. The complete UV-Vis absorption spectrum of the IrPS complex is shown in Fig. S1 of the Electronic Supplementary Information (ESI). 
\begin{figure}[h]
 \centering
 \includegraphics[height=5cm]{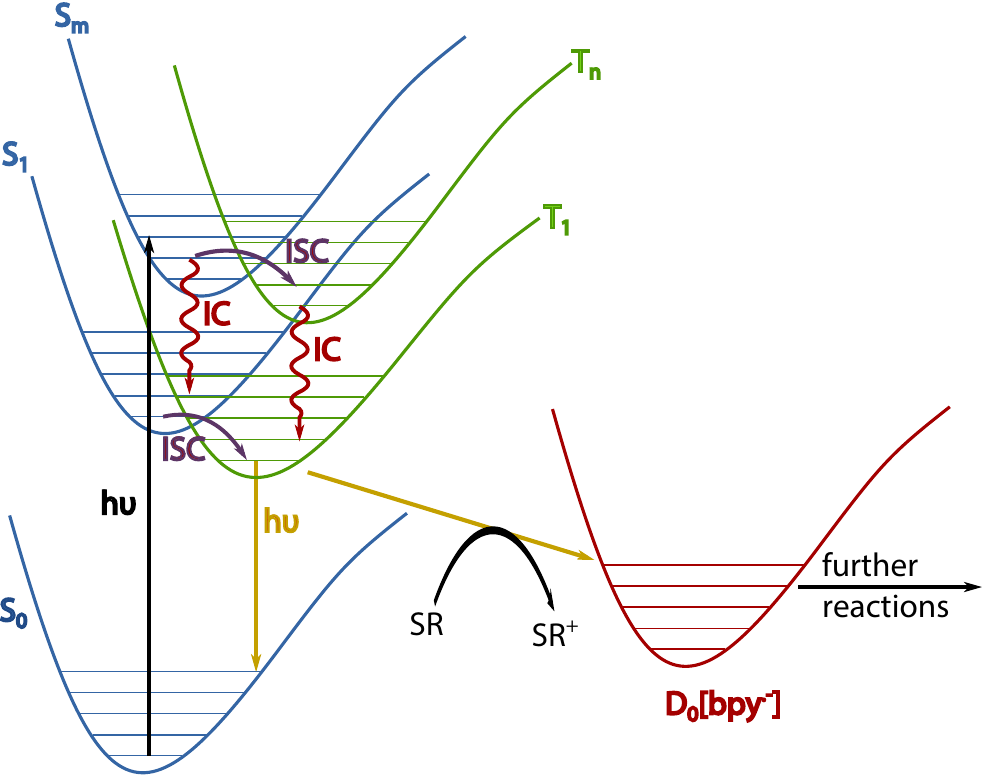}
 \caption{Potential energy surface scheme of the relevant potentials as well as the photophysical and -chemical reactions of the [Ir(ppy)$_2$(bpy)]$^+$, adapted from Ref. \citenum{Bokarev2015}. In the present study we focus on the dynamics of the reactive T$_1$ state and in particular the electron transfers involving the PS.}
 \label{fig_pes}
\end{figure}
The photophysical and -chemical processes in the Ir-based PS complex, which take place upon the excitation into the singlet MLCT band at 355 nm are summarized in Fig. \ref{fig_pes}. 
The photoexcitation of the molecule occurs within the Franck-Condon regime from the singlet ground state S$_0$ to a singlet MLCT state labeled S$_m$. The lowest excited bright state corresponds to a $\rm \pi d_{x^2-y^2} \rightarrow \pi^\ast(ppy)$ electronic transition.~\cite{Bokarev2012,Bokarev2014}
Excitation is followed by an ultrafast relaxation via internal conversion (IC) and intersystem crossing (ISC) into the lowest triplet MLCT state of $\rm \pi d_{x^2-y^2} \rightarrow \pi^\ast(bpy)$ character labeled T$_1$. 
Recently, ISC rates for similar Ir(III)-complexes were measured using femtosecond optical transient absorption spectroscopy \cite{Hedley2009} and TR photoluminescence measurements.\cite{Hedley2009,Pomarico2016}
It was found that the ISC and relaxation into T$_1$ state occurs within 70 - 100 fs. 
This triplet state has a long lifetime, which can be chemically tuned from some hundreds of nanoseconds to several microseconds. \cite{Gartner2012}
Once in the long-lived T$_1$ state, the IrPS can be reduced to a doublet D$_0$ state by the SR and the excess electron residing on the $\rm \pi^\ast(bpy)$ orbital~\cite{Bokarev2014} can be transferred from the PS$^{-}$ complex to the WRC, as shown in step III in Fig. \ref{fig_reactioncycle}. 
As a competing mechanism to the reduction, a radiative or non-radiative deactivation of the T$_1$ back into the ground state can occur. 
The photoluminescence lifetimes of the T$_1$ state in tetrahydrofuran (THF) solution both with and without the SR (present as a co-solvent in a THF/TEA mixture with a 5/1 volume ratio) were measured to be 13 ns and 370 ns, respectively.\cite{Neubauer2014} 
A complete study of the underlying photoinduced dynamics and more importantly the mechanistic aspects governing these processes at the PS and the WRC sites is often severely hindered when using only optical methods, since they are not sensitive to the oxidation state of the central metal ion and the molecular geometric structure. On the contrary, employing ultrafast X-ray spectroscopies allows accessing in an element-specific manner the active sites, i.e. the different metal ions, and to capture the transient electronic and geometric structure changes occurring during the photocatalytic function of the system.
\newline
The present study is motivated by early pump-probe X-ray absorption spectroscopy (XAS) studies, in which the oxidation states\cite{Gawelda2006, Chen2007} and local geometric structures\cite{Gawelda2006, Chen2007, Vanko2015} of liquid-phase molecular systems have been investigated. Later, TR XAS has been used to investigate the triplet excited state of the Ir model system \ce{Ir(ppy)3} \cite{Goeries2014} and the combination with DFT allowed the study of this excited state of a pure solvated Ir photosensitizers \cite{Smolentsev2018}. Picosecond- and femtosecond-resolved X-ray studies of a bimetallic donor-acceptor Ru-Co complex (a prototypical photocatalytic system) have shown how ultrashort hard X-ray spectroscopies can be efficiently used to follow both the intra- and intermolecular charge transfer processes at the optically silent sites of photocatalysts.~\cite{Canton2013, Canton2015}. In a similar study, pump-probe XAS, paired with all-optical experiments, was used to investigate intramolecular electron transfer processes in a whole family of homo- and heterodinuclear Cu(I)/Ru(II) complexes
~\cite{Hayes2018}. The applicability of high-resolution energy detected high-energy resolution fluorescence detection (HERFD) \cite{Hamalainen1991} XAS as well as X-ray Emission Spectroscopy (XES) to answer questions in catalysis has been further demonstrated using Fe complexes \cite{Bauer2014}. In very recent experiments, the structural and electronic configurations of a Co \cite{Moonshiram2016b} and Ni\cite{Moonshiram2016} proton reduction catalyst were characterized.
\newline
By a similar approach we investigate here the primary electron transfer step that experiences the IrPS \ce{[Ir$^{\textnormal{\tiny{III}}}$(ppy)2(bpy)]+} in the above introduced photocatalytic system after optical excitation. The goal is to characterize the electronic configuration of the IrPS prior and after the electron transfer event and to identify the involved orbitals in order to obtain a validated microscopic picture of this crucial step.


\section{Experimental Measurements}
Time-resolved spectra of the X-ray absorption near edge structure (XANES) of \ce{[Ir(ppy)2(bpy)]+} were collected at the 7ID-D beamline of the Advanced Photon Source (APS) both in total fluorescence yield (TFY) as well as in the HERFD mode. More details concerning the experimental setup and the pump-probe techniques available at this beamline can be found in Refs \cite{March2011, Haldrup2012, Vanko2015}. Here we will only briefly summarize the experimental conditions present during the pump-probe measurements, which allowed us to record high quality X-ray absorption and X-ray emission spectra of both the ground and excited states of the PS complex. 
The APS in 24 bunch mode delivered an average X-ray flux of 4$\times$10$^{11}$ photons/s at the Ir L$_3$-edge (11.2 keV) in pulses with a repetition rate of 6.5 MHz. The X-ray probe pulses were monochromatized to $\Delta E / E = 5\times$10$^{-5}$ and focused to a spot of 4$\times$5 $\mu$m$^2$ on the sample consisting of the sample solution in a 100 $\mu$m thick free flowing liquid jet. The optical pump laser repetition rate was set to 931 kHz, i.e. the 7th sub-harmonic of the X-rays. The inter-pulse spacing of 1.07 $\mu$s allows for a full ground state recovery between consecutive laser pulses. The laser wavelength was frequency tripled to 355 nm and laser pulses with 142 nJ energy were used to excite the sample. The laser focus was set to 13$\times$21 $\mu$m$^2$, thus slightly larger than the X-ray spot size, leading to a laser peak intensity of 6.6 GW/cm$^2$.
TFY XAS were measured using a scintillation detector and simultaneously HERFD XAS were acquired to obtain XAS with sub core-hole lifetime limited energy resolution. The HERFD XAS were measured with a Johann\cite{Johann1931} spectrometer employing a Ge(800) analyzer crystal to selectively detect photons at the L$\alpha_1$ peak energy of 9175 eV.
Three different sample solutions were measured: i) 15 mM of the pure PS in \ce{CH3CN}, ii) 15 mM of the PS in a 1:4 mixture of TEA and \ce{CH3CN} and iii) 12.5 mM PS and 9 mM \ce{Fe3(CO)12} in a 1:5 mixture of TEA and \ce{CH3CN}.
Here, \ce{CH3CN} is chosen as solvent since it provides a higher solubility for the PS than THF, which is used in the optimized catalytic system. It was previously shown that the photocatalytic system works also with \ce{CH3CN} as solvent, however, with lower yields.\cite{gartner11_6998} The high PS concentration is necessary due to the thin sample thickness.\newline
Ultrafast optical absorption studies as well as time-resolved luminescence measurements were already performed on the PS dissolved in acetonitrile.\cite{Tschierlei2016} They showed that also in this Ir complex the MLCT state is populated within the first 100 fs and the lifetime of its luminescence amounts to 60 ns. Since these experiments were carried out with air-saturated samples it was advantageous for a comparison on common ground to use also in the present experiments air-saturated solutions.
\section{Theoretical Calculations}
X-ray spectra originating from the ground singlet $S_0$ and first excited triplet $T_1$ electronic states were calculated in the dipole approximation at the first principles restricted active space self-consistent field (RASSCF) level and perturbative LS-coupling scheme for the spin-orbit coupling \cite{Malmqvist2002,bokarev19_}. 
This method, due to its correlated nature, is prerequisite to describe  X-ray absorption resulting from excited states (here T$_1$).
A relativistic ANO-RCC basis set \cite{Roos2004,Roos2005} of triple-zeta quality for iridium and its first coordination shell and of double-zeta quality for other atoms was employed. For the detailed description of the protocol, see e.g. Ref. \citenum{Bokarev2015b}. 
The complex has a low $C_2$ point group symmetry.
Thus, we prefer not to use octahedral symmetry labels.
The active space comprised three Ir 2p, two pairs of highly correlated bonding $\sigma$d  and antibonding $\sigma$*d orbitals, the $\pi \rm d_{x^2-y^2}$ orbital, two mainly non-bonding $ \rm d_{xz}$ and $\rm d_{yz}$ orbitals, and one $\pi^\ast$(bpy) orbital (see Fig.~\ref{fig_rasscf}b)). This allows for the description of the S$_0$ and T$_1$ states (originating from $\pi\rm d_{x^2-y^2} \rightarrow \pi^\ast(bpy)$ excitation) as well as of the Ir L-edge core excited states. 
This setup should provide reliable results for transition metal compounds, where static electron correlation could play an important role.\cite{Bokarev2012} Spectra were calculated at the equilibrium geometries of the S$_0$ and T$_1$ states taken from Ref.~\citenum{Bokarev2012}. 
Molecular orbital energies were obtained using long-range corrected density functional theory (LC-BLYP with a LANL2DZ/6-31G(d) basis set)\cite{bokareva15_1700} since RASSCF does not provide energies of canonical orbitals.
\section{Results and Discussion}
\begin{figure}[h]
 \centering
 \includegraphics[width=\columnwidth]{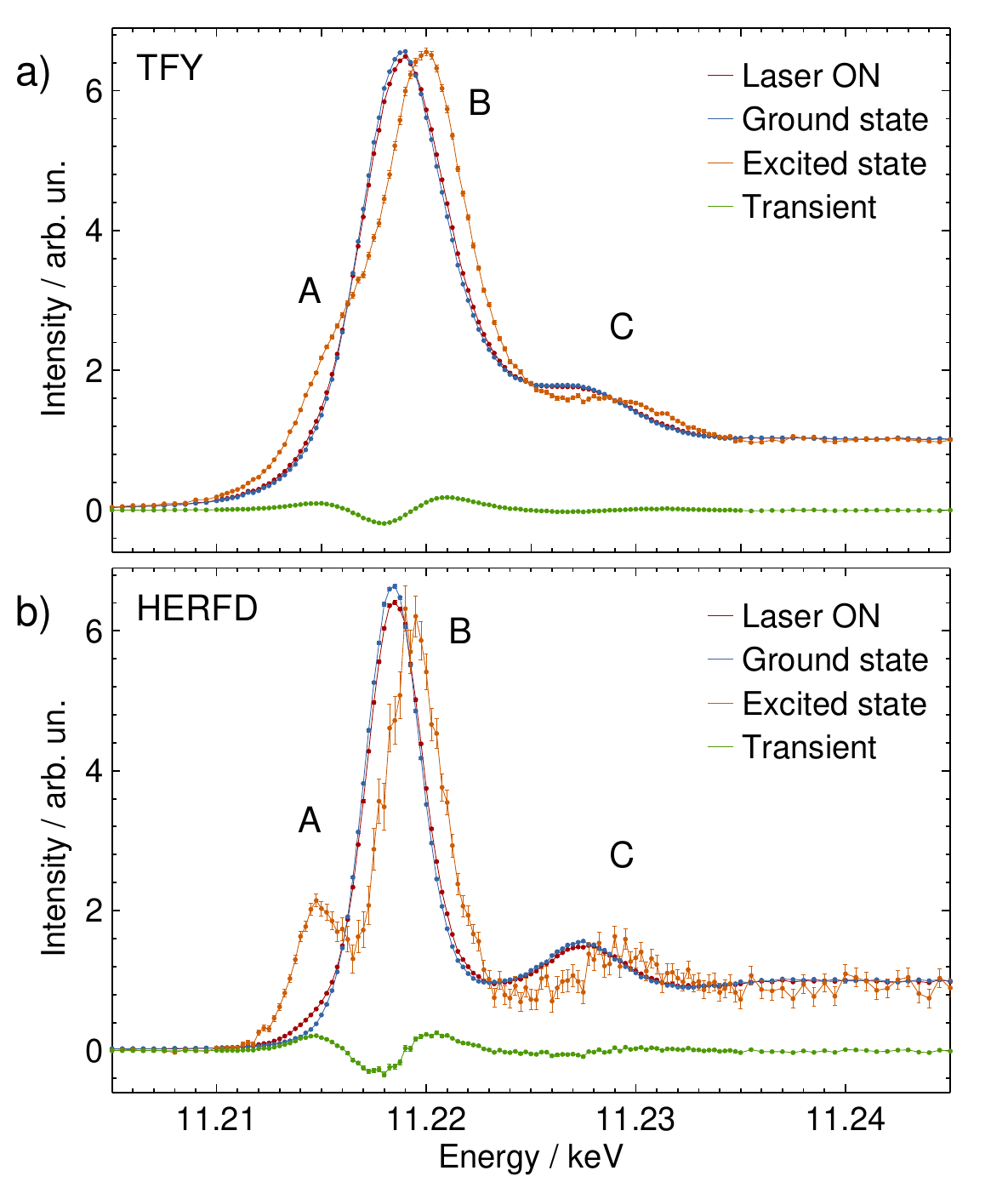}
 \caption{L$_3$-edge XANES spectra measured in a) TFY and b) HERFD of the laser excited ensemble (red), the ground state (blue), the reconstructed excited state (orange) and the transient difference (green) of Ir[(ppy)$_2$(bpy)]$^+$ ~in CH$_3$CN.}
 \label{fig_xanesl}
\end{figure}

Figure \ref{fig_xanesl} shows the TFY and HERFD XANES spectra of the bare PS in \ce{CH3CN} (i.e. in absence of TEA and WRC) with and without laser excitation, as well as the transient difference and the reconstructed excited state spectrum. 
The spectrum of the laser excited sample (Laser ON) was determined to contain an excited state fraction of 12\%, see ESI for details on the determination of the excited state fraction. This is used to reconstruct the excited state (ES) spectrum by removing the remaining (dominant) contribution of the ground state (GS) from the laser excited spectrum.

\begin{figure}[h]
\centering
  \includegraphics[width=\columnwidth]{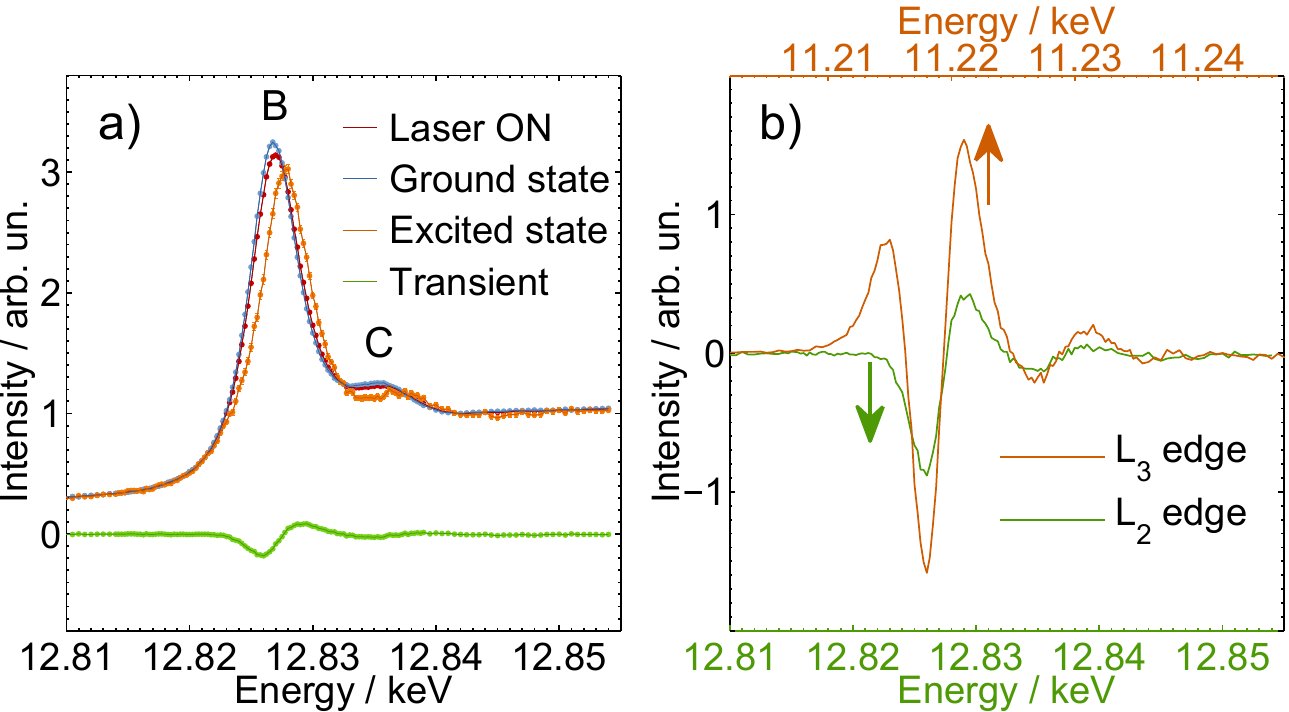}
  \caption{a) TFY L$_2$-edge XANES of the laser excited ensemble (red), the reconstructed
excited state (orange), the ground state (blue) and the transient difference (green) of
 \ce{[Ir(ppy)2(bpy)]+} in \ce{CH3CN}. In b), a comparison between this L$_2$-edge transient (green)
to the L$_3$-edge transient (orange) is shown, the transients are normalized to the respective
edge jump and scaled to 100 \% excited state fraction.} 
  \label{fig_l2edge_on_off}
\end{figure}
\begin{figure}[h!]
 \centering
 \includegraphics[width=\columnwidth]{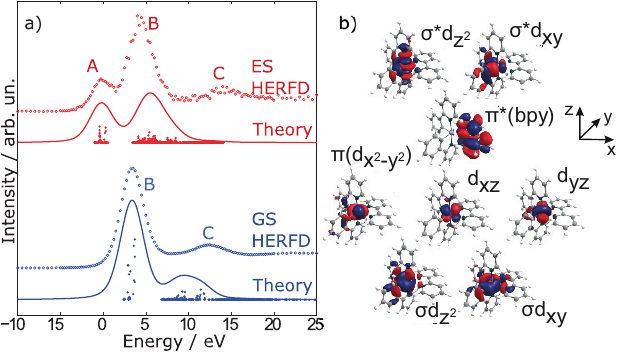}
 \caption{Simulated L$_3$-edge XANES of the S$_0$ GS and the T$_1$ MLCT ES (a), the x-axis is given relative to the 2p$_{3/2}$ ionization energy of 11215 eV. The orbital active space (apart from 2p orbitals) used for the RASSCF calculations is shown in b).}
 \label{fig_rasscf}
\end{figure}
\begin{figure}[h!]
 \centering
 \includegraphics[width=\columnwidth]{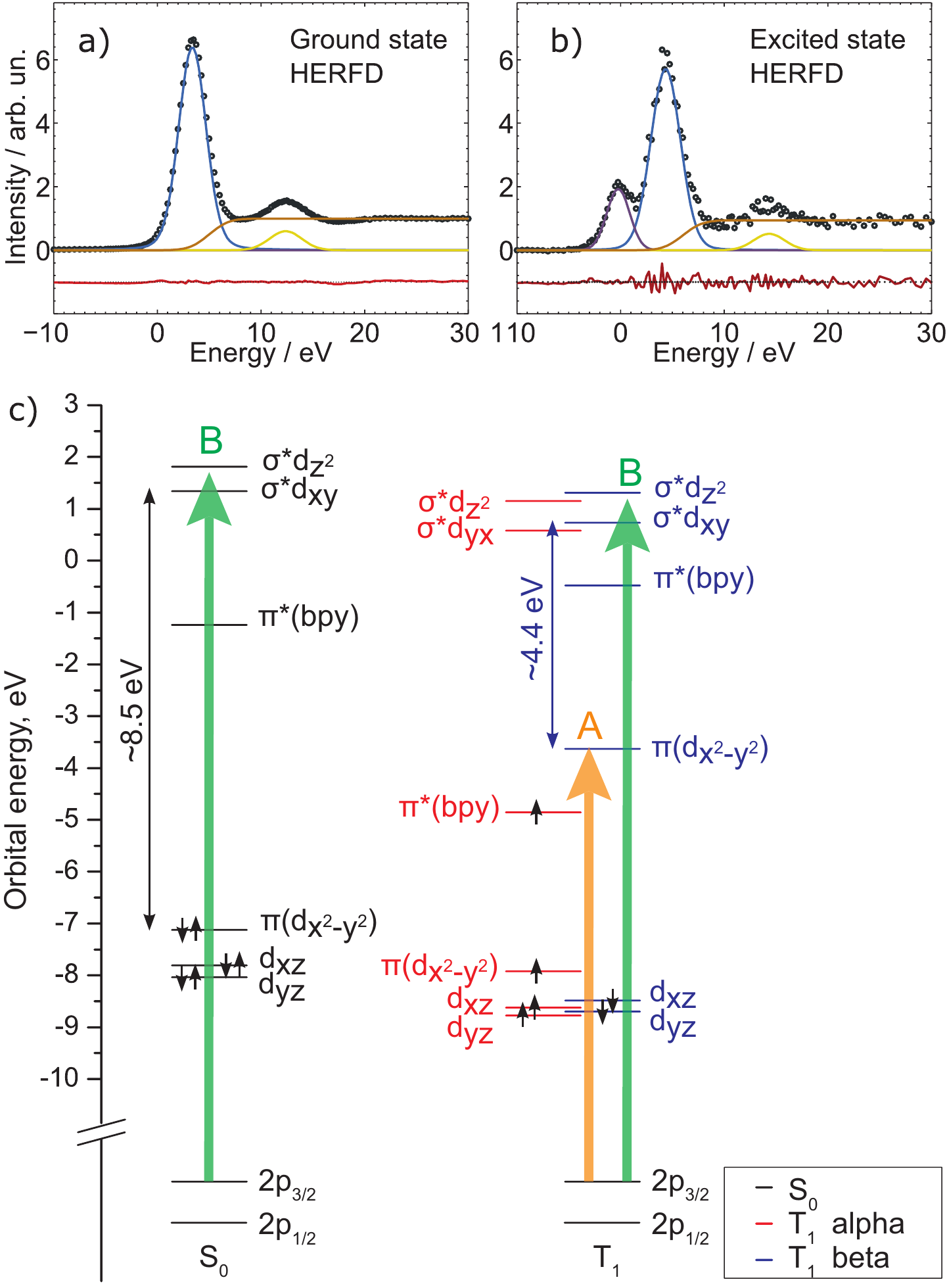}
 \caption{Reconstructed HERFD XANES spectra together with measured data of [Ir(ppy)$_2$(bpy)]$^+$ ~in CH$_3$CN in its GS a) and ES b). The x-axis is given relative to the 2p$_{3/2}$ ionization energy of 11215 eV. 
 c) Diagram showing the relevant orbitals as computed with restricted (S$_0$) and unrestricted (T$_1$) density functional theory.
 The orbital manifold of the T$_1$ state contains two subsets: $\alpha$ orbitals occupied by spin-up electrons (red) and $\beta$ ones with spin-down electrons (blue) which have different energies. The transitions of the L$_3$-edge for the GS involve the orbitals 2p$_{3/2}$, $\sigma^\ast \rm d_{z^2}$, and $\sigma^\ast \rm d_{xy}$ (green arrow, band B) and for the ES additionally transition to spin-down $\rm \pi d_{x^2-y^2}$ (orange arrow, band A). 
 The L$_2$-edge XANES contains the corresponding transitions starting from the 2p$_{1/2}$ sub-shell (not shown with arrows).}
 \label{fig_transitions}
\end{figure}
 \begin{table}[h]
 \tiny
   \caption{Upper panel: Energies of L$_3$-edge XANES features, all energies are given in eV relative to the 2p$_{3/2}$ ionization energy of 11215 eV. Lower panel: Energies of L$_2$-edge XANES features, all energies are given in eV relative to the 2p$_{1/2}$ ionization energy of 12824 eV.}
   \label{table_transitions}
	\begin{tabular*}{0.48\textwidth}{@{\extracolsep{\fill}}lllll}
     \hline
 & A & B & C & edge \\
     \hline
Ground state ($L_3$)   &  & 3.8$\pm$0.1 & 12.8$\pm$0.3& 5.4$\pm$1.6\\
Excited state  ($L_3$)  & 0.2$\pm$0.2&4.8$\pm$0.3 & 14.7$\pm$0.4&6.4$\pm$ 8.8\\
Shift GS-ES ($L_3$) & & +1.0$\pm$0.4 & +1.9$\pm$0.5& +1.0$\pm$ 9.0\\
     \hline

Ground state ($L_2$)   &  & 2.7$\pm$0.1 & 11.7$\pm$0.1& 4.5$\pm$1.1\\
Excited state  ($L_2$)  & -0.8$\pm$0.4 & 3.7$\pm$0.2 & 13.4$\pm$0.3&5.7$\pm$ 3.3\\
Shift GS-ES ($L_2$) & & +1.0$\pm$0.3 & +1.7$\pm$0.4& +1.2$\pm$ 3.5\\
   \hline
   \end{tabular*}
\end{table}
\begin{figure}[h!]
 \centering
 \includegraphics[height=7.5cm]{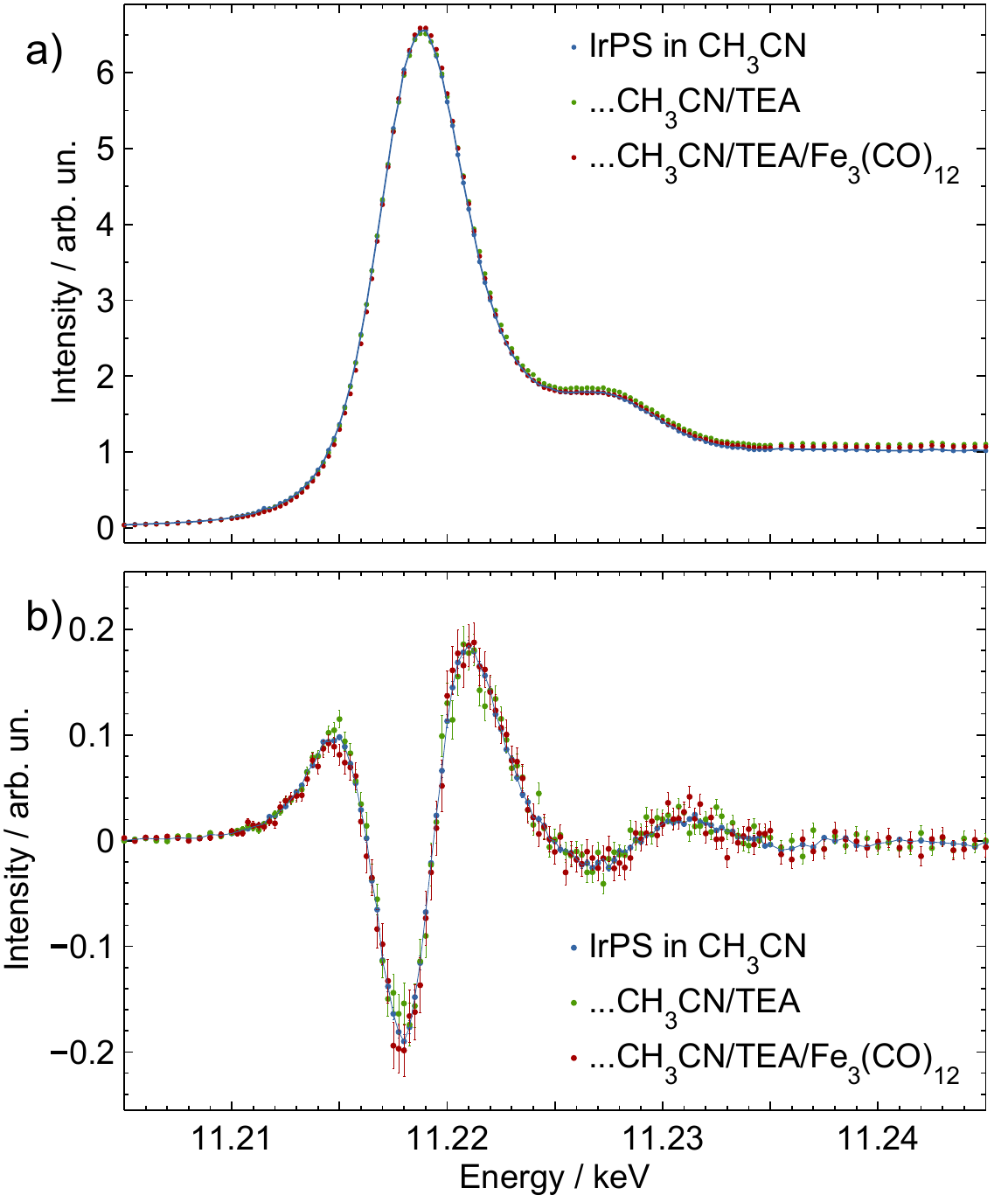}
 \caption{Comparison of XANES of [Ir(ppy)$_2$(bpy)]$^+$ in CH$_3$CN (blue), as well as in CH$_3$CN/TEA (green) and in CH$_3$CN/TEA/Fe$_3$(CO)$_{12}$(red) mixtures, respectively. In a) the ground state spectra and in b) the transient differences 100~ps after excitation are shown. The static spectra and the transient differences are arbitrarily scaled in intensity to simplify comparison.}
 \label{fig_solvmixesl}
\end{figure}
\begin{figure}[h!]
\centering
\includegraphics[width=\columnwidth]{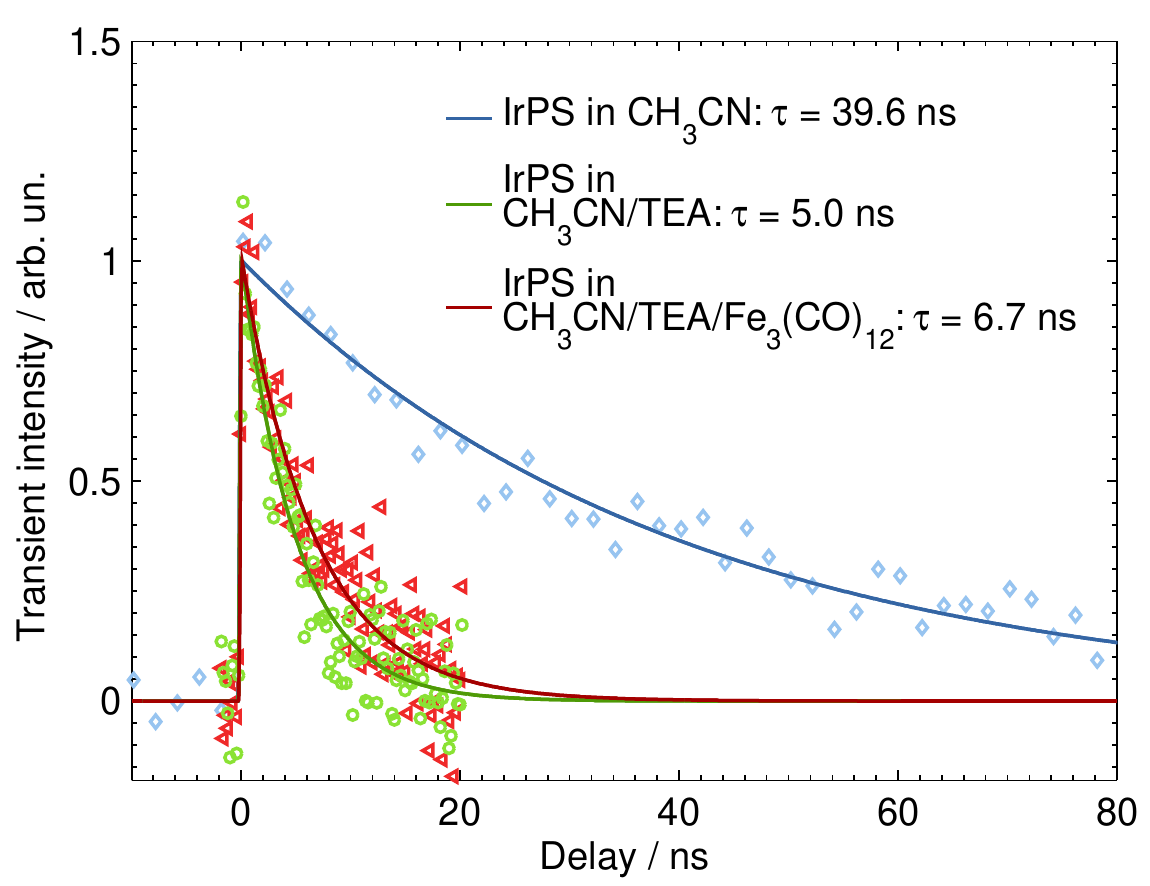}
\caption{Time delay scan of Ir[(ppy)$_2$(bpy)]$^+$ ~in CH$_3$CN (blue), as well as in CH$_3$CN/TEA (green) and CH$_3$CN/TEA/Fe$_3$(CO)$_{12}$ (red) mixtures. A fit of an exponential decay resulted in excited state lifetimes of (39.6$\pm$2.8) ns, (5.0$\pm$0.6) ns and \mbox{(6.7$\pm$0.7) ns} for the three samples, respectively.}
\label{fig_delayscansl}
\end{figure}
\noindent The features of the TFY spectra are extremely broad due to the 5.25 eV L$_3$-edge core-hole lifetime broadening \cite{Krause1979a}. The HERFD spectra show the same features as the TFY spectra, however, with a by about 2 eV improved resolution of 3.0-3.5 eV. Thus for the further analysis we focus on the HERFD spectra in Fig. \ref{fig_xanesl}b). The prominent features of the spectra are labeled A, B and C, cf. Tab. \ref{table_transitions}. In the GS spectrum a very intense white line feature B around 11.219 keV and a second transition C at higher energy of about 11.229 keV are visible. The ES spectrum contains an additional peak A at a lower energy of about 11.215 keV, which is not visible in the GS spectra. Furthermore, B and C shift about 1 eV - 2 eV to higher energies in the ES compared to the GS, thus evidencing the oxidation of the Ir center from Ir$^{\rm III}$ to Ir$^{\rm IV}$.\newline
The L$_2$-edge measured in TFY is presented in Fig. \ref{fig_l2edge_on_off}a) and exhibits the same features in the GS as the L$_3$-edge. However, in the ES a clear feature A is missing. The absence of this feature becomes even clearer when comparing the transient differences of the GS and the ES (see Fig. \ref{fig_l2edge_on_off}b)), in the L$_2$-edge the first positive peak of the L$_3$-edge is not present. 
\newline
The assignment of the L$_3$-edge features is made on the basis of RASSCF calculations. 
The comparison of the experimental and theoretical spectra for both ground and excited states can be found in Fig. \ref{fig_rasscf}a).
The spectrum consists of several tens of transitions, with some of them having small intensity summing up to a notable feature due to their number. 
The assignment of multi-reference many-body RASSCF wave functions is quite involved due to their multiconfigurational character, which is additionally complicated by spin-orbit coupling.
Here, for simplicity, the information is reduced to transitions between single orbitals (Fig. \ref{fig_rasscf}b)), having however non-integer occupations. An orbital energy scheme for the relevant transitions discussed in the following is given in Fig. \ref{fig_transitions}c).
\newline
The GS spectrum contains two features denoted as B and C in Fig. \ref{fig_rasscf}a) (blue line).
Peak B mainly consists of $\rm 2p \rightarrow \sigma^\ast d_{z^2}$ and $\rm 2p \rightarrow \sigma^\ast d_{xy}$ transitions. 
Remarkably, some of the transitions having non-vanishing intensity are of $\rm 2p \rightarrow \pi^\ast(bpy)$ nature and can be considered as charge-transfer excitations.
Peak C in turn is due to two-electron shake up transitions, where in the region of 7-10\,eV $\rm 2p \rightarrow \sigma^\ast d_{z^2}$ transitions are occurring simultaneously with excitations from $\rm \pi d_{x^2-y^2}$ and $\rm d_{xz}$ orbitals to $\rm \pi^\ast(bpy)$ ones.
Transitions in the 10-13\,eV range are very similar but in this case $\rm \sigma^\ast d_{xy}$ is accepting the electron instead of $\rm \sigma^\ast d_{z^2}$.
\newline
Features in the T$_1$ spectrum have similar nature as those of the ground state spectrum with the only difference that the triplet configuration has a hole in the $\rm \pi d_{x^2-y^2}$ orbital and an additional electron in the $\rm \pi^\ast(bpy)$ orbital.
This fact gives rise to an additional feature A in the excited state spectrum where transitions correspond mainly to $\rm 2p \rightarrow \pi d_{x^2-y^2}$ excitation. 
-This assignment is in agreement with that reported for the higher symmetric homoleptic ${[\textnormal{Ru(bpy)3}]^{2+}}$ complex. \cite{Gawelda2006}
The main type of transitions contributing to feature B is of $ \rm 2p \rightarrow \sigma^\ast d_{z^2}$ and $\rm 2p \rightarrow \sigma^\ast d_{xy}$ nature.
In addition to shake ups mentioned above, one sees also contributions from $\rm d_{xz,yz} \rightarrow \pi d_{x^2-y^2}$ and $\rm \pi^\ast(bpy) \rightarrow \sigma^\ast d_{xy}$ transitions because of the additional hole/electron present in the T$_1$ state.
Interestingly, the low-energy flanks of the bands A and B in the T$_1$ spectrum have mainly spin-forbidden character gaining intensity due to spin-orbit coupling. 
This means that the initial state is mostly triplet, while the final ones are predominantly singlet.
\newline
To further quantify the exact energy positions of the XAS features, the HERFD XANES spectra have been fitted using a superposition of an arctan-broadened absorption edge and Voigt profiles for the discrete transitions. The corresponding bands of the fit are shown in Fig. \ref{fig_transitions}a) and b) and the peak and edge position obtained from this fit can be found in Tab. \ref{table_transitions}. The same fitting procedure was also applied to the L$_2$-edge TFY XAS (see ESI). The obtained energies of the respective features are added to the table. 
The energy shift between the A and B transitions reflects the energy difference between the $\rm \pi d_{x^2-y^2}$ and the two $\sigma^\ast$ orbitals of (4.6$\pm$0.4) eV which is within the error bars also confirmed by the respective features of the L$_2$-edge.
The splitting is in good accord with the theoretical value reported in Figs. \ref{fig_rasscf}a) and \ref{fig_transitions}c).
 Most importantly for the application of the time-resolved XAS measurements to the catalytic hydrogen production process, we are able to determine the occupation of the 5d-orbital and to monitor the optically induced charge separation with concomitant creation of a vacancy in the 5d-shell. This should in the next step allow for monitoring the arrival of an additional charge from the SR to the PS*.

The static TFY XANES spectra of the PS in \ce{CH3CN}, of the PS in the TEA/\ce{CH3CN} mixture and 
of the PS in the mixture together with the WRC \ce{Fe3(CO)12} are indistinguishable within our experimental accuracy, see Fig. \ref{fig_solvmixesl}. This allows us to conclude that the addition of the SR as well as of the WRC does not severely affect the geometric or electronic structure of the PS in the GS. 
The transient TFY XANES of all three samples after optical excitation are qualitatively similar as well, thus 100 ps after photoexcitation the PS is presumably in the same electronic state independent of the addition of the SR and the WRC.\newline
Time delay scans of the excited state of these three samples have been taken at the energy of maximum transient intensity, i.e. at 11221.5 eV (see Fig. \ref{fig_delayscansl}). A fit of a convolution of a Gaussian broadened step function with an exponential decay delivered MLCT excited state lifetimes of (39.6$\pm$2.8) ns, (5.0$\pm$0.6) ns and (6.7$\pm$0.7) ns for the PS in \ce{CH3CN}, the IrPS in the TEA/\ce{CH3CN} mixture and the PS together with the \ce{Fe3(CO)12} in the TEA/\ce{CH3CN} mixture, respectively. The ground state recovery time of (39.6$\pm$2.8) ns is similar to the previously reported luminescence decay time of 60 ns which is supposed to reflect the depopulation of the MLCT state into the ground state.\cite{Tschierlei2016}
The lifetime shortening associated with the addition of TEA by a factor of 8
 is due to the electron transfer (ET) from TEA to the IrPS.
 This is clearly observed by the appearance of the A feature upon optical excitation and the associated depopulation of the $\rm \pi d_{x^2-y^2}$-orbital
 and its subsequent disappearance reflecting the refilling of this orbital by the ET process. Previous optical experiments were not able to monitor this refill process and thus the final configuration of the MLCT deactivation step.~\cite{Neubauer2014}
The slightly longer excited state lifetime of 7 ns in the fully functionally photocatalytic system can be attributed to the lower TEA concentration. Using the ES lifetime $\tau_0$ without and $\tau$ with the TEA present we can calculate the TEA concentration $c_{\textnormal{\tiny{TEA}}}$ dependent quenching rate constant $k_q$ of the ES via ET using the Stern-Volmer equation \cite{Neubauer2014}
\begin{equation}
k_q(c_{\textnormal{\tiny{TEA}}})= \frac{1}{\tau} - \frac{1}{\tau_0} = 1.76 \times 10^{8} \textnormal{ s}^{-1} \:.
\end{equation}
With the TEA concentration $c_{\textnormal{\tiny{TEA}}} = 1.44$ M this results in a bimolecular quenching respectively ET rate of $k_{\textnormal{\tiny{ET}}} = 1.22 \times 10^{8}$ (Ms)$^{-1}$. The ET rate for a diffusion limited reaction is estimated to $2.06 \times 10^{10}$ (Ms)$^{-1}$ (see ESI for details). According to this, only about every 170th encounter of a TEA molecule with an excited PS leads to an electron transfer process.
\newline
The results are well in line with previous findings obtained by time-resolved photoluminescence measurements on the PS. In that case quenching of the sensitizer phosphorescence by TEA was studied in tetrahydrofuran solutions.\cite{Neubauer2014} A bimolecular quenching rate of 5.9$\times$10$^7$ (Ms)$^{-1}$ was observed while the diffusion controlled collision rate was estimated to 1.4$\times$10$^{10}$ (Ms)$^{-1}$ indicating that only one out of 200 collisions results in a quenching event. This is in good agreement with the ET efficiency found in the present experiments. In the previous study also ab initio calculations on the collision complex were performed.\cite{Neubauer2014} They indicate that only at specific collision geometries an ET is possible and only a small fraction of the collisions occurs with such a geometry. This can explain the small number of encounters resulting in ET. The advantage of the present measurements is that not only depopulation of a luminescent state is observed but the filling of the hole at the Ir atom by an electron from the SR proving the proposed ET step.

\section{Conclusions}
We have applied picosecond-resolved X-ray spectroscopies to investigate an Ir-based PS for photocatalytic hydrogen generation. Thereby functionally relevant bimolecular electron transfer steps in a homogeneous solution have been monitored for the first time.
We
were able to quantify the electronic structural changes of the PS as they take place during the photocatalytic cycle by mapping the (metal centered) unoccupied orbitals of the Ir PS in ground and excited states using mainly L$_3$-edge XANES. The features in the TFY spectra are extremely broadened due to the >5 eV core-hole lifetime and applying HERFD XANES enabled an improvement in resolution of about 2 eV down to 3.0 eV - 3.5 eV. 
This allowed for an accurate analysis of the XANES features and decomposing the edge into its main transitions, i.e. to accurately determine their energy positions and relative intensities. The detailed analysis was guided by high-level ab initio calculations of ground and excited state spectra.
Equipped with this knowledge about relevant X-ray fingerprints, the Ir PS in the fully functioning photocatalytic system was investigated. Here, we were able to directly observe the bimolecular electron transfer process from TEA to the Ir PS via a reduction of the MLCT excited state lifetime of the Ir PS by a factor of about 8. 
More specifically, we can monitor the oxidation state of the Ir atom. After optical excitation and trapping in the MLCT state, the Ir$^{\rm III}$ ion is oxidized to Ir$^{\rm IV}$ leading to the appearance of a new band in the X-ray absorption spectrum.  Upon bimolecular electron transfer from the sacrificial reductant the hole on the Ir 5D shell is refilled, yielding back Ir$^{\rm III}$ and its characteristic spectrum. Due to the element specifity the extra electron on the bpy ligand does not yield noticeable changes in the Ir L-edge spectrum. Hence, the present setup is not sensitive to the electron relay from the Ir PS to the WRC. Therefore, in future studies 
 it would be desirable to also apply time-resolved X-ray absorption and emission at the Fe-edges of the  WRC (\ce{Fe3(CO)12}) to monitor also the last steps (III and IV) of the reaction cycle.
 \newline
 To summarize, the present proof-of-principle study has enhanced our understanding of functionally relevant bimolecular electron transfer reactions and thus will pave the road to rational optimization of the performance of homogeneous photocatalytic systems.
\section*{Conflicts of Interest}
There are no conflicts to declare.
\section*{Acknowledgements} We thank Chris Milne and Jakub Szlachetko for their help with setting up the HERFD XAS measurements and for the loan of analyzer crystals, Frank de Groot and Pieter Glatzel for discussions regarding the interpretation of the HERFD XAS measurements as well as Daniel Haskel for the loan of the Ir(IV) reference.\newline
This work is financed by the European XFEL, by the Deutsche Forschungsgemeinschaft (DFG) via SFB 925 (project A4), and by the Hamburg Centre for Ultrafast Imaging (CUI). AB acknowledges support from the International Max Planck Research School for Ultrafast Imaging and Structural Dynamics (IMPRS-UFAST), and we acknowledge the European Cluster of Advanced Laser Light Sources (EUCALL) within work packages PUCCA (CB), which has received funding from the European Union's Horizon 2020 research and innovation programme under grant agreement No 654220. WG acknowledges the National Science Centre (NCN) in Poland for grant SONATA BIS 6 No. 2016/22/E/ST4/00543. ZN, {\'E}GB and GV  from the Wigner 'Lend{\"u}let' (Momentum) Femtosecond Spectroscopy Research Group were supported by the ELKH under contract LP2013-2/2020 Program of the Hungarian Academy of Sciences (LP2013-59), the Government of Hungary and the European Regional Development Fund under grant VEKOP-2.3.2-16-2017-00015, and the National Research,
Development and Innovation Fund (NKFIH) under contract FK124460. SIB and OK acknowledge financial support from DFG (grants BO 4915/1-1 and KU 952/10-1).\newline
G.D, A.M.M. and S.H.S. acknowledge support from the U.S. Department of Energy, Office of Basic Energy Sciences, Division of Chemical Sciences, Geosciences, and Biosciences through Argonne National Laboratory.
This research used resources of the Advanced Photon Source, a U.S. Department of Energy (DOE) Office of Science User Facility operated for the DOE Office of Science by Argonne National Laboratory under Contract No. DE-AC02-06CH11357.
\section*{Author Contributions}
WG, CB, OK, SL and AB designed the research; NR and HJ synthesized the sample; NR, HJ and MB contributed to the conception; GD, AMM and SHS built the high-repetition rate pump-probe X-ray spectroscopy setup; AB, TA, EGB, ZN, GV, GD, AMM and SS performed the X-ray experiments; SIB performed the simulations; AB analyzed the experimental data; AB, SIB and WG wrote the manuscript with contributions from all co-authors.


\end{document}